\documentclass[12pt]{article}
\usepackage{amsmath, amsthm, amscd, amsfonts, amssymb, graphicx, color}
\usepackage[bookmarksnumbered, colorlinks, plainpages]{hyperref}
\hypersetup{colorlinks=true,linkcolor=red, anchorcolor=green, citecolor=red, urlcolor=black, filecolor=black, pdftoolbar=true}
\newtheorem{theorem}{Theorem}[section]

\theoremstyle{definition}
\newtheorem{definition}[theorem]{Definition}

\theoremstyle{remark}
\newtheorem{remark}{Remark}[section]
\numberwithin{equation}{section}

\begin{document}

	\setcounter{page}{1}
	
	
	\begin{center}
		{\Large  \bf{Impact of symmetry inheritance  on  conformally flat spacetime. }}
		
		\bigskip
		{\bf Musavvir Ali,$^{1,a}$ Mohammad Salman,$^{1,b,\ast}$   and Sezgin Altay Demirbag$^{2,c}$}		
		
		$^{1}$ Department of Mathematics, 
		
		Aligarh Muslim University, Aligarh-202002, India.
		
		$^{2}$ Department of Mathematics, 
		
		Istanbul Technical University,  Istanbul-34469, Turkey.
		
		\footnote{Email addresses:  $^{a}$\url{musavvir.alig@gmail.com},    $^{b}$\url{salman199114@gmail.com} ($^{*}$Corresponding Author), and $^{c}$\url{saltay@itu.edu.tr}}
		
		\textbf{}
	\end{center}
	
	\bigskip
	
	\begin{abstract}
		The goal of this research paper is to investigate curvature inheritance symmetry in conformally flat spacetime. Curvature inheritance symmetry in conformally flat spacetime is shown to be a conformal motion. We have proven that a conformally flat spacetime reduces to Einstein spacetime if admits curvature inheritance symmetry. A few results on conformally flat spacetimes that obey Einstein's field equation with or without a cosmological constant, if  admits the curvature inheritance symmetry. The energy-momentum tensor is to be covariantly constant in a 4-dimensional relativistic perfect fluid spacetime which is also conformally flat spacetime, admits curvature inheritance, and obeys Einstein's field equations in the presence of a cosmological constant. Moreover, it is also obtained that such spacetimes with perfect fluid satisfy the the vacuum-like equation of state consecutively it is dark matter.  Finally, in the third part of the article, the case compatible with all Theorems from Theorem \ref{Th2.1} to Theorem \ref{Th2.5n}  is shown. On the other hand, it has also been emphasized that it is an example of de Sitter spacetime. It has been demonstrated that this spacetime also has a conformal killing vector.
	\end{abstract}
	
	\noindent{\small {\bf  Mathematics Subject Classification:} 53C15; 53C25;  53B20;  53C80; 83C05.\\
		{\bf Keywords:} Curvature  inheritance symmetry, Conformal curvature tensor,  conformally  flat spacetime, Lambda-CDM model.}
	
	\section{Introduction and Preliminaries}\label{sec1}
	
	This study examines whether (non-trivial) curvature inheritance symmetry occurs in conformally flat spacetime. Several conformally flat spacetimes undoubtedly admit these symmetries in all circumstances. In spacetime $V_4$ the curvature tensor associated to Lorentzian metric $g$ of signature $(-, +,+,+)$ through the Levi-Civita connection, denoted by $R^h_{ijk}$, the components of Ricci tensor $R_{ij}  = R^h_{ihj}$ and the Ricci scalar are given by $R = g^{ij} R_{ij}.$ The symbols $(;)$,   $(,)$ and $\pounds_\xi$ represent the covariant, partial, and Lie derivatives, respectively. The standard symmetrization and skew-symmetrization are indicated by round and square brackets respectively. An n-dimensional  Riemannian manifold $(V_{n},g)$ is said to be flat if its Riemann curvature tensor is zero  everywhere; otherwise, it is called non-flat.

	The curvature tensor is a significant concept in differential geometry (\cite{Besse}, \cite{Eishart}, \cite{UC de}, \cite{Abdussattar and B Dwivedi}) and, more specifically in general relativity theory (\cite{Zengin1}-\cite{ZAhsan3}). As a result, the study of its symmetries, namely the inheritance of curvature, is a significant subject that has recently attracted a lot of attention. This paper is a further attempt at symmetries admitted by the conformally flat spacetime, and we assure that it will present an effective contribution to the study of symmetries. However, we find rare examples on the topic, but this study will impact in a reasonably elegant fashion.
	
	The Bianchi identities explain the gravitational field interaction with matter and free gravitational components, shown by the curvature tensor in the general theory of relativity. The primary goal of all research in gravitational physics is to build a gravitational potential that satisfies Einstein's field equations (EFEs) with a cosmological constant. According to \cite{KarmerD}, the interaction between matter and gravitation through EFEs with a cosmological term is given by 
	\begin{equation}\label{2.1}
		R_{ij} - \dfrac{R}{2} g_{ij} + \wedge g_{ij} =  k ~  T_{ij},
	\end{equation}
	
	\noindent  where $\wedge$, $R_{ij}$,  $ g_{ij}$,   $T _{ij}$,  and  $k$  denotes the cosmological term, Ricci tensor,  metric,   energy-momentum tensor, and constant respectively. Expression of the energy-momentum tensor   in  perfect fluid spacetime   as follows,
	\begin{equation}\label{2.2}
		T_{ij}=(\mu + p) u_{i}u_{j}+ p g_{ij},
	\end{equation} 
	where  $p$ represents  the isotropic pressure,  $\mu$  represents the energy density,  and $u^{i}$ is the four velocity vector   of the flow satisfying  $g_{ij} u^{j} = u_{i}$  for all  $i$  and $u_{i}u^{i}= -1$.

	The objective is accomplished for a chosen matter distribution when we put symmetry assumptions on the geometry that are compatible with the dynamics of the selected matter distribution. The following equation represents these geometrical symmetries of spacetime\cite{yanoK}
	\begin{equation}\label{2.3}
		\pounds_\xi  A =2  \varOmega A,
	\end{equation}
	where $ \pounds_\xi $  denotes the Lie derivative along the vector field $\xi^{i}$. \textquoteleft A\textquoteright  ~~  signifies a geometrical/physical quantity, and $\varOmega$ is a scalar function. The vector field $\xi^{i}$ can be time-like  ($\xi^{i}\xi_{i}<0$),  space-like  ($\xi^{i}\xi_{i}>0$), or null ($\xi^{i}\xi_{i}=0$).
	
	The metric inheritance symmetry or conformal motion (Conf M) \cite{Katzin} along a conformal Killing vector (CKV) $\xi^{i} $ for A = $g_{ij}$ in equation\eqref{2.3} is a straightforward example that can be given here 
	\begin{equation}\label{2.4}
		\pounds_\xi g_{ij}   =  2 \psi  g_{ij}, 
	\end{equation} 
	
	\noindent where $\psi$ represents the conformal function and the most fundamental symmetry on $(V_4, g) $ is motion (M), which is obtained by setting $ A = g_{ij} $ and $\varOmega = 0$ in equation \eqref{2.3}, and $\xi^{i}$ is referred to as a killing vector \cite{yanoK}. In literature, there are more than 30 different geometric symmetries. (For a detailed study of symmetry inheritance, see  \cite{Duggal}-  \cite{Ahsan}).
	
	In 1992,  K.L. Duggal  \cite{Duggal} introduced the concept of curvature inheritance (CI), which is the generalization of curvature collineation (CC), and many other collineations were defined by Katzin in 1969 \cite{Katzin} and also by Duggal \cite{klduggal2}. Most recently, Ali et al. (\cite {AliM}-\cite{AliM3}) investigated symmetry inheritance for conharmonic curvature tensor, and curvature inheritance in Ricci flat spacetimes and A. A. Shaikh et al. have studied  CI in \cite{Sheikh}  for M-projectively flat spacetime. Also, lots of authors have studied the  curvature symmetry in the context of general relativity (\cite{Zengin1}, \cite{Zahsan12},  \cite{TupperHall}-\cite{Hallshabbir}). In 2004, G.S. Hall published a book on the field of  symmetries, and curvature structure in general relativity \cite{hall}. Furthermore,  Abdussatar and B. Dwivedi extend the work of symmetries   of spacetime for conharmonic curvature tensor \cite{Abdussattar and B Dwivedi}.   
	\begin{definition}
		\cite{Duggal} A spacetime $(V_4, g)$ admits curvature inheritance symmetry along a 	 smooth vector $\xi^{i}$, such that 
		\begin{equation} \label{2.5}
			\pounds_\xi R^h_{ijk} = 2 \varOmega R^h_{ijk}, 
		\end{equation}
		where $\varOmega = \varOmega(x^{i})$ represents inheriting factor or inheritance function.
	\end{definition}
	\noindent If  $\varOmega = 0$, then $ 	\pounds_\xi R^h_{ijk} = 0$ and $\xi$ is said to follow a symmetry known as curvature collineation (CC) admitted by $V_4$.
	The CI equation \eqref{2.5}	 can also be written as
	\begin{equation}\label{2.6}
		R ^h_{ijk; l} \xi^l - R^l _{ijk} \xi^h_{;l} + R^h_{ljk} \xi ^l_{;i} +R^h_{ilk} \xi ^l_{;j} +R^h_{ijl} \xi ^l_{;k} = 2 \varOmega  R^h_{ijk}.
	\end{equation}
	\begin{definition}
		\cite{Duggal}	A spacetime $(V_4, ~g)$ admits Ricci inheritance (RI) symmetry along a 	 smooth vector  $\xi^{i}$,   such that
		\begin{equation} \label{2.7}
			\pounds_\xi R_{ij} = 2 \varOmega R_{ij}.
		\end{equation}
	\end{definition}
	\noindent Contraction of equation  \eqref{2.5} implies equation \eqref{2.7}. Thus,  every curvature inheritance is  Ricci inheritance symmetry (i.e., CI $\Rightarrow$ RI), but the converse may not hold. If $\varOmega =0$, then RI reduces to RC; otherwise, if $\varOmega \neq 0$, then called proper RI.
	\cite{Weyl}	The conformal curvature tensor is expressed as follows       
	\begin{equation}\label{2.8}
		C^h_{ijk}= R^h_{ijk}+\dfrac{1}{2}(\delta^h_j R_{ik}-\delta^h_k R_{ij}+ R^h_j g_{ik}- R^h_k g_{ij})+\dfrac{R}{6}
		(g_{ij}\delta^h_k- g_{ik}\delta^h_j).
	\end{equation}
	\begin{definition}
		A Riemannian space is conformally flat  \cite{Weyl} iff 
		\begin{equation}\label{2.9}
			C^h_{ijk}=0,  ~~~~  (n>3).
		\end{equation}
	\end{definition}    
	\begin{definition}
		(\cite{Besse})	$\bf	Einstein ~ Spacetime.$
		If the Ricci  and metric tensors are proportional so that, in coordinates
		\begin{equation}
			R_{ij} = \mu g_{ij},
		\end{equation}
		for some function $\mu$, then  $\mu = \dfrac{R}{4}$  and $V_4$ is called an Einstein spacetime (and so if $V_4$ has constant curvature it is an Einstein spacetime).
	\end{definition} 
	The plan of the present paper is as follows: After Section 1 introduction and preliminaries, we study curvature inheritance in a conformally flat spacetime. Section 2 deals with conformally flat spacetime admitting curvature inheritance symmetry, and here also, we have discussed some properties of such spacetimes. We obtain some interesting results about the conformally flat perfect fluid spacetimes admitting curvature inheritance symmetry and obeying EFEs with the cosmological term. Moreover, we prove that the pressure and density of these spacetimes are constant. We obtain a result for purely electromagnetic distribution. Section 3,  the case compatible with all Theorems from Theorem \ref{Th2.1} to Theorem \ref{Th2.5n} is shown.
	Finally, we gives brief discussion in the Section 4.
	
	\section{Main Results}\label{sec2}
	
	In a conformally flat spacetime the conformal curvature tensor is given by 
	\begin{equation}\label{3.1}
		R^l_{ijk} = - \dfrac{1}{2}(\delta^l_j R_{ik}-\delta^l_k R_{ij}+ R^l_j g_{ik}- R^l_k g_{ij}) - \dfrac{R}{6}
		(g_{ij}\delta^l_k- g_{ik}\delta^l_j). 	
	\end{equation}
	A necessary conditions for a spacetime $(V_4, g)$ to admit  curvature inheritance symmetry is derived in (\cite{Duggal}, equation (2.2)) and may be expressed as,  the following identity satisfy  by the curvature tensor,
	\begin{equation}\label{2.2n}
		R^a_{ijk} g_{al}+ R^a_{ljk} g_{ai} = 0. 
	\end{equation}
	
	\noindent Taking the Lie derivative of \eqref{2.2n}	and using \eqref{2.5} and \eqref{3.3}, we obtain
	\begin{equation}\label{3.2}
		R^a_{ijk}\hbar_{al}+ R^a_{ljk}\hbar_{ai}=0, 
	\end{equation}
	here
	\begin{equation}\label{3.3}
		\hbar_{ij} = \pounds_\xi g_{ij} ~~ \mbox{and}~~  h = \hbar_{ij} g^{ij}.
	\end{equation}
	Let us assume that $P_{lijk} = R^a_{ijk}\hbar_{al}+ R^a_{ljk}\hbar_{ai}$ = 0.	In special case that $V_4$ is a conformally flat spacetime, we substitute $R^l_{ijk}$ as given by \eqref{3.1} into \eqref{3.2} and we get 
	\begin{eqnarray}\label{3.4}
		P_{lijk} &=&\hbar_{lj} R_{ik} - \hbar_{lk} R_{ij} + \hbar_{ij} R_{lk} - \hbar_{ik} R_{lj} - g_{lj} S_{ik} + g_{lk} S_{ij}\notag - g_{ij} S_{lk}\\
		&& + g_{ik} S_{lj} +  \dfrac{R}{3} (g_{lj} \hbar_{ik} - g_{lk} \hbar_{ij} + g_{ij} \hbar_{lk}- g_{ik} \hbar_{lj} ) = 0,
	\end{eqnarray}
	
	\noindent where $S_{ij} = \hbar_{ik} R^k_{j}$  and $S = g^{ij} S_{ij}$. We multiply \eqref{3.4} by $g^{lj}$ and sum to obtain
	\begin{equation}\label{3.5}
		3 S_{ij} + 	S_{ji} = S g_{ij} + \dfrac{R}{3}\hbar_{ij} + h (R_{ij} - \dfrac{R}{3} g_{ij}).
	\end{equation} 
	In \eqref{3.5} we interchange the indices i and j and subtract  the resulting equation from \eqref{3.5}, we find $S_{ij} = S_{ji}.$  Thus, equation \eqref{3.5} reduces to 
	\begin{equation}\label{3.6}
		S_{ij} = \dfrac{1}{4} [S g_{ij} + \dfrac{R}{3}\hbar_{ij} + h (R_{ij} - \dfrac{R}{3} g_{ij})].
	\end{equation}
	Next, let  $P'_{lijk} = P_{lijk} + P_{jkli} = 0$, substituting $P_{lijk}$ from \eqref{3.4} then using $S_{ij}$ as given by \eqref{3.6}, we thus obtain 
	\begin{equation}\label{3.7}
		P'_{lijk} = E_{ik} P_{lj} - P_{ik} E_{lj} = 0,
	\end{equation}
	where $P_{ij} = \hbar_{ij} - \dfrac{h}{4} g_{ij}$~and~$ E_{ij} = R_{ij} - \dfrac{R}{4}g_{ij}$.
	If $E_{ij}= 0$, the conformally flat space is an Einstein space. If a conformally flat spacetime is also an Einstein spacetime such that $R=0$, then the spacetime is flat. Hence, we rule out the possibility of having a conformally flat spacetime as an Einstein spacetime with $R=0$ and consider this outcome. From  (\cite{Duggal}, Theorem 7), we have that every proper CI in an Einstein space with  $(R\neq0)$ is a proper conf M with conformal function. This article considers spacetime that is conformally flat and has non-zero scalar curvature. Thus, we shall assume that spacetime is conformally flat but not Einstein spacetime.
	
	We proceed by choosing $E_{ij}\neq 0$ in \eqref{3.7}, $P_{ij} = \lambda E_{ij}$, and here we consider $\lambda$ an arbitrary scalar. Substitution of $P_{ij}$ and $E_{ij}$ in equation $P_{ij} = \lambda E_{ij}$, leads to 
	\begin{equation}\label{3.8}
		\hbar_{ij} = \pounds_\xi g_{ij} = 2 \phi g_{ij} + \lambda R_{ij},~~~~~~~~\mbox{where~~~ $2\phi = \dfrac{1}{4}(h - \lambda R)$.}
	\end{equation}

	\par  Let  $(V_4, g)$ denotes a conformally flat spacetime admitting curvature inheritance symmetry. Next we substitute $S_{ij}$ and $S$ in equation \eqref{3.6}. In the resulting equation we use \eqref{3.3} and  \eqref{3.8} with $\lambda\neq 0$ to substitute $\hbar_{ij}$ and $h$ and find that a $(V_4)$ must satisfy the condition 
	\begin{equation}\label{3.9}
		R^{k}_{j} R_{ik} =  \alpha g_{ij} + \beta R_{ij},
	\end{equation}
	where
	\begin{equation}
		\alpha = \dfrac{1}{4}[R'- \dfrac{R^{2}}{3}],~~~~~~~~~ \beta = \dfrac{R}{3},~~~~~~~~~~ R' = R^{i}_{j} R^{j}_{i}.
	\end{equation}
	
	Now, taking Lie derivatives of equation \eqref{3.1} and use of equations \eqref{2.5}, \eqref{2.7} and \eqref{3.3}, leads to
	\begin{equation}
		\pounds_\xi R^l_{ijk} = \pounds_\xi [ - \dfrac{1}{2}(\delta^l_j R_{ik}-\delta^l_k R_{ij}+ R^l_j g_{ik}- R^l_k g_{ij})- \dfrac{R}{6}
		(g_{ij}\delta^l_k- g_{ik}\delta^l_j)],~~~~~~~~~~~~~~~~~
	\end{equation}
	\begin{eqnarray}
		\pounds_\xi R^l_{ijk} &=&  - \dfrac{1}{2}[\delta^l_j \pounds_\xi R_{ik}-\delta^l_k \pounds_\xi R_{ij}+ (\pounds_\xi R^l_j) g_{ik} + R^l_j (\pounds_\xi g_{ik}) - (\pounds_\xi R^l_k) g_{ij} - R^l_k (\pounds_\xi g_{ij}) ]\notag\\ 
		&-&  \dfrac{1}{6}[(\pounds_\xi R)
		(g_{ij}\delta^l_k- g_{ik}\delta^l_j) + R 	( \delta^l_k \pounds_\xi g_{ij}    - \delta^l_j  \pounds_\xi g_{ik})],
	\end{eqnarray}
	\begin{eqnarray*}
		\pounds_\xi R^l_{ijk} &=&  - \dfrac{1}{2}[2 \varOmega (\delta^l_j  R_{ik}-\delta^l_k R_{ij})+ (2\varOmega R^l_j - R^x_{j} \hbar ^l_{x}) g_{ik} + R^l_j \hbar_{ik} - (2 \varOmega  R^l_k - R^y_{k} \hbar^l_{y}) g_{ij}\notag\\  
		&-& R^l_k \hbar_{ij} ] -  \dfrac{1}{6}[(2\varOmega R - R')
		(g_{ij}\delta^l_k- g_{ik}\delta^l_j) + R 	( \delta^l_k \hbar_{ij}    - \delta^l_j  \hbar_{ik})],
	\end{eqnarray*}
	\begin{eqnarray}
		\pounds_\xi R^l_{ijk} &=& 2 \varOmega [ - \dfrac{1}{2}(\delta^l_j R_{ik}-\delta^l_k R_{ij}+ R^l_j g_{ik}- R^l_k g_{ij})- \dfrac{R}{6}
		(g_{ij}\delta^l_k- g_{ik}\delta^l_j)] +
		\dfrac{1}{2}(- R^x_{j} \hbar ^l_{x} g_{ik} \notag\\ &-& R^y_{k} \hbar^l_{y} g_{ij} - R^l_k \hbar_{ij}) -  \dfrac{1}{6}[( - R')
		(g_{ij}\delta^l_k- g_{ik}\delta^l_j) + R 	( \delta^l_k \hbar_{ij}    - \delta^l_j  \hbar_{ik})].
	\end{eqnarray}
	From equation \eqref{3.1}, we observe that \eqref{3.2} reduces to
	\begin{eqnarray}
		\pounds_\xi R^l_{ijk} &=&  \pounds_\xi R^l_{ijk} +
		\dfrac{1}{2}	(- R^x_{j} \hbar ^l_{x} g_{ik} - R^y_{k} \hbar^l_{y} g_{ij} - R^l_k \hbar_{ij}) \notag\\ &-&  \dfrac{1}{6}[( - R')
		(g_{ij}\delta^l_k- g_{ik}\delta^l_j) + R ( \delta^l_k \hbar_{ij}    - \delta^l_j  \hbar_{ik})],
	\end{eqnarray}
	\begin{eqnarray}
		3( R^x_{j} \hbar ^l_{x} g_{ik} + R^y_{k} \hbar^l_{y} g_{ij} + R^l_k \hbar_{ij}) + ( - R')
		(g_{ij}\delta^l_k- g_{ik}\delta^l_j) \notag\\
		+ R 	( \delta^l_k \hbar_{ij}    - \delta^l_j  \hbar_{ik}) = 0,
	\end{eqnarray}
	\begin{eqnarray}
		3 [2\phi (R^l_{j} g_{ik} +  2 R^l_{k} g_{ij}) + \lambda ( R^x_{j}  R^l_{x} g_{ik} +  R^y_{k}  R^l_{y} g_{ij} + R^l_{k} R_{ij}) ]  \notag\\ - \lambda R_{ij} R^{ij} (\delta ^l_{k} g_{ij}- \delta ^l_{j} g_{ik} ) +
		\lambda R (\delta ^l_{k}R_{ij}- \delta ^l_{j} R_{ik}) =0.
	\end{eqnarray}
	Contraction on $l$ and $j$ in above equation gives that 
	\begin{eqnarray}\label{3.18}
		2 [ \phi R +  \lambda R^{2} + \lambda  \alpha  ] g_{ik} + [ 4 \phi + 2 \lambda  \beta - \lambda R] R_{ik} = 0,
	\end{eqnarray}
	\begin{eqnarray}\label{2.19}
		R_{ik} =  - \dfrac{2( \phi R +  \lambda R^{2} + \lambda  \alpha ) }{4 \phi + 2 \lambda \beta - \lambda R} g_{ik},
	\end{eqnarray}
	where $R^2 = R^x_{y} R^y_{x} =  R_{xy} R^{xy}$. If multiplying  equation \eqref{2.19} by $ g^{ik}$ we have 
	\begin{equation}\label{2.20}
		- \dfrac{2( \phi R +  \lambda R^{2} + \lambda \alpha ) }{4 \phi + 2 \lambda \beta - \lambda R} = \dfrac{R}{4}.	
	\end{equation}
	By using equation \eqref{2.19} and \eqref{2.20} we get
	\begin{equation}\label{2.21}
		R_{ik} = \frac{R}{4} g_{ik}.
	\end{equation}

	Since $E_{ik} = R_{ik} - \frac{R}{4} g_{ik} = 0$  and  $P_{ik} = \lambda E_{ik}$ we find
	\begin{equation}\label{2.22n}
		\hbar_{ik} = \pounds_\xi g_{ik} = \frac{h}{4} g_{ik}, ~~~~~~~(h= 8\varOmega).
	\end{equation}
	In equation \eqref{2.21}, $R\neq0$, otherwise this spacetime is flat. We summaries the above analysis by stating, 
	\begin{theorem}\label{Th2.1}
		Every  curvature inheritance symmetry	in a conformally flat spacetime is a   proper conformal motion.
	\end{theorem}
	Theorem \ref{Th2.1} can also be proved by another method. \noindent Theorem 7 in \cite{Duggal} \textquotedblleft Every proper CI in an Einstein space $R\neq 0$ is a proper conformal motion  with conformal function $\psi$." By using the  Theorem 7 in \cite{Duggal} and equation \eqref{2.21}, we get the Theorem \ref{Th2.1}.
	From \eqref{2.21}, we have another result,
	\begin{theorem}\label{Th2.2}
		A conformally flat spacetime reduces to an Einstein spacetime if it admits curvature inheritance symmetry along a vector field $\xi$.
	\end{theorem}

	\noindent Riemannian manifold with harmonic curvature
	\begin{equation}
		(X,Z) (\nabla_{X} S) (Y,Z) -  (\nabla_{Y} S)  = (div R) (X,Y,Z),
	\end{equation}
	here $R(X,Y,Z)$  and  $S(X,Y)$ denotes Riemannian curvature tensor and Ricci tensor respectively. 
	\begin{definition}
		\cite{Derdz} If $(div R) (X, Y, Z) = 0$,  an algebraic curvature tensor R is considered harmonic. If a Riemannian manifold M curvature tensor field R is harmonic, the manifold is said to be R-harmonic.
	\end{definition}
	Since,
	\begin{equation}
		\nabla_{h} R^h_{ijk} =  \nabla _{k} R_{ij}  - \nabla _{j} R_{ik}. 
	\end{equation}
	\noindent From Theorem \ref{Th2.2}, we have $R_{ij} = \frac{R}{4} g_{ij} $. Thus,  we  have 
	\begin{equation}
		\nabla_{h} R^h_{ijk} = \nabla _{k}(\frac{R}{4} g_{ij}) - \nabla _{j}(\frac{R}{4}  g_{ik}) = 0, ~~~~~~~~~~~~~\mbox{(Since R is constant)}.
	\end{equation}
	This implies that spacetime is R-harmonic. Thus, we have 
	\begin{theorem}\label{Th2.4n}
		If a  conformally flat spacetime admits curvature inheritance then  this  spacetime has  harmonic curvature.
	\end{theorem}
	\begin{theorem}\label{Th2.5n}
		If a conformally flat spacetime admits CI then this spacetime is of constant curvature.
	\end{theorem}
	\begin{proof}
		From Theorem \ref{Th2.2}, conformally flat spacetime is an  Einstein spacetime. Remembering that this spacetime  is a conformally flat, the proof is clear.
	\end{proof}
	\begin{remark}
		In relativity, spaces with constant curvature are significant. Cosmology is extremely familiar with the importance of spaces with constant curvature. The most basic cosmological model assumes that the universe is isotropic and homogeneous. According to the cosmological principle, it is impossible to distinguish between two places in the universe because all spatial directions are equal. The Robertson-Walker metrics are cosmological solutions to the Einstein field equations with a three-dimensional space-like surface with constant curvature. In contrast, the de Sitter universe model has a four-dimensional space with a constant curvature \cite{Nalikar}.
	\end{remark}
	\noindent

	The matter inheritance, which is determined by the well-known symmetry of the energy momentum tensor,
	\begin{equation}\label{3.25N}
		\pounds_\xi T_{ij} = 2 \varOmega T_{ij}.
	\end{equation}
	
	\begin{theorem}\label{Th2.4}
		If a conformally flat spacetime admits CI and obeys the Einstein's filed equations with cosmological constant, then the spacetime admits matter inheritance property along a vector $\xi$. 
	\end{theorem}
	
	\begin{proof}
		In view of \eqref{2.21}, equation \eqref{2.1} yields
		\begin{equation}\label{3.25}
			( \wedge - \dfrac{R}{4}) g_{ij} = \kappa T_{ij}.
		\end{equation}
		Operating the Lie derivative on equation \eqref{3.25}, we have 
		\begin{equation}\label{3.26}
			(\wedge - \dfrac{R}{4}) \pounds_\xi g_{ij} = \kappa \pounds_\xi T_{ij}.
		\end{equation}
		From 	Theorem \ref{Th2.1}, $\xi$ is a conformal Killing vector field,  we have
		\begin{equation}\label{3.27}
			\pounds_\xi g_{ij} = 2 \varOmega g_{ij}.
		\end{equation}
		Using equations \eqref{3.26} and \eqref{3.27}, we write
		\begin{equation}\label{3.28}
			\pounds_\xi T_{ij} = 2 \varOmega T_{ij}.
		\end{equation}
		The proof is completed.
	\end{proof}
	
	\begin{theorem}\label{Th2.8}
		If a conformally flat  spacetime admits CI  and obeys the Einstein's filed equations with cosmological constant, then dark matter exists there.
	\end{theorem}
	\begin{proof}
		From \eqref{2.2} we have,
		\begin{equation}\label{2.40}
			T = 3p -\mu.
		\end{equation}
		Multiplying equation  \eqref{3.25} by $g_{ij}$ we get
		\begin{equation}\label{2.41n}
			\kappa	T =  4 \wedge - R.
		\end{equation}
		By using the equation \eqref{2.40} and \eqref{2.41n} we obtain
		\begin{equation}\label{2.42n}
			R = 4 \wedge - (3p-\mu)\kappa. 
		\end{equation}
		By using equations \eqref{2.1}  and Theorem \ref{Th2.2} we find
		\begin{equation}\label{2.34n}
			(\wedge - \frac{R}{4}) g_{ij} = \kappa T _{ij}.
		\end{equation}
		Multiplying the above equation by $u^{i}$ and $u^{j}$ and using equation \eqref{2.2} gives 
		\begin{equation}\label{2.44}
			R = 4 \wedge + 4 k \mu.
		\end{equation}
		From equation \eqref{2.42n} and \eqref{2.44} we obtain 
		\begin{equation}\label{2.45}
			\mu + p = 0.
		\end{equation}
		Since $ \mu + p = 0 $, spacetime admitting CI with Einstein's field equations is dark matter. Thus, the proof is completed. 
	\end{proof}
	If  $  p + \mu =0$,  the scalar curvature is equivalent to cosmological constant \cite{KarmerD}. According to \cite{RH1},  R. H. Ojha  termed this as Phantom Barrier. In 1981 A. Guth  proposed the topic of cosmic inflation \cite{Guth} and explained the similar conditions for the same in the universe. The  authors explained the term inflation in their paper \cite{Luca} as rapid expansion of the spacetime that might occurred just after the $Bing ~~Bang$. 
	
	Also, we conclude that equation \eqref{2.45} represents the vacuum-like equation of state. Thus, we have next theorem,
	\begin{theorem}
		If a conformally flat spacetime admits CI  and obeys the EFEs with cosmological constant, then this spacetime satisfies the vacuum-like equation of state.
	\end{theorem}
	
	\begin{theorem}
		If a conformally flat spacetime admits CI and obeys Einstein's field equation with the cosmological constant, it is an  $\wedge$CDM (Lambda cold dark matter)  model.
	\end{theorem}
	\begin{proof}
		A very simple   relationship between energy density $\mu$ and  the isotropic pressure $p$   of the perfect fluid is $p = \omega \mu $, where $\omega$  is the equation of state parameter. Experimentally, the equation of state parameter is connected to the evolution of the energy density and the universe's expansion. In \cite{Saini}, we obtained $\omega = - 1$, which corresponds to the vacuum energy, using equation \eqref{2.45}, which is the $\wedge$CDM (Lambda cold dark matter)  model or Lambda-CDM model.
	\end{proof}

	By using equation \eqref{2.42n} and \eqref{2.45} we get 
	\begin{equation}\label{2.37n}
		p = \frac{1}{\kappa} (\wedge - \frac{R}{4}),~~~~~~~~~~~~~ \mu = \frac{1}{\kappa}(\frac{R}{4}- \wedge).
	\end{equation}
	Consequently, we can summarize the result as follows. 
	\begin{theorem}\label{Th2.8n}
		Energy density $\mu$ and isotropic pressure $p$ are constants if a conformally flat spacetime admits CI and obeys the EFEs with the cosmological constant.
	\end{theorem}

	From  equation \eqref{2.34n} and Theorem \ref{Th2.2} we have, 
	
	\begin{theorem}
		The energy momentum tensor is covariantly constant in a conformally flat spacetime admits CI that obeys Einstein's field equation with the cosmological constant.
	\end{theorem}

	\begin{theorem} \label{Th2.9}
		If a conformally flat spacetime satisfies the EFEs without cosmological constant, admits CI along  $\xi$ and satisfies 
		\begin{equation}\label{3.38n}
			\varOmega_{;ij} = \frac{\varOmega}{2}(\frac{R g_{ij}}{3} - 2  R_{ij}),
		\end{equation}
		then it admits curvature collineation.
	\end{theorem}
	\begin{proof}
		From \cite{Duggal},  we have  
		\begin{equation}\label{3.5nn}
			\pounds_\xi \mu = -2 \square\varOmega   -2 \varOmega  \mu -2 \varOmega _{;ij}, u^{i} u^{j},
		\end{equation} 
		where $\square$ is the Laplacian operator defined as $\square \varOmega  = \varOmega_{;ij} g^{ij}$. By using Theorem \ref{Th2.2} and equation \eqref{3.38n} we get 
		\begin{equation}\label{2.41}
			\varOmega_{;ij}=-\frac{1}{12}\varOmega R g_{ij}.
		\end{equation}
		From equation \eqref{2.41} we have 
		\begin{equation}\label{2.42}
			\square \varOmega = - \frac{1}{3} \varOmega R.
		\end{equation}
		Using equations \eqref{3.5nn} - \eqref{2.42}, we obtain
		\begin{equation}\label{2.43}
			\pounds_\xi \mu  =  \frac{\varOmega}{2}  (R- 4\mu).	
		\end{equation}
		Now using  Theorem \ref{Th2.8n}, we get either $\varOmega =  0$ or $R = 4\mu$. Form equation \eqref{2.37n} and if $p =0$ ($\kappa \neq 1$) gives $\mu$ = 0,  $R = 0$.  Since this is conformally flat and Einstein, it reduces to flat. Therefore $ p \neq 0$. The spacetime admits curvature collineation (CC). The proof is completed.
	\end{proof}
	\begin{theorem}\label{Th2.10}
		If a conformally flat spacetime satisfies the EFEs without  cosmological constant,  admits  CI along a vector field $\xi$ and satisfies equation \eqref{3.38n} then CI reduces to motion.
	\end{theorem}
	\begin{proof}
		From equation \eqref{2.22n}, every curvature inheritance in a conformally flat spacetime gives $\pounds_\xi g_{ij} = 2 \varOmega g_{ij}. $ From Theorem \ref{Th2.8n} the proof is clear.
	\end{proof}
	
	Theorem \ref{Th2.10} can be also proved by another method, Theorem 9 in \cite{Duggal} \textquotedblleft Every CC in an Einstein space ( R $\neq$ 0, n $ >$ 2) is a motion.\textquotedblright
	~ By  using Theorem 9 in \cite{Duggal}, Theorem \ref{Th2.2} and Theorem \ref{Th2.9} we get Theorem \ref{Th2.10}. 
	
	\begin{theorem}
		A general relativistic conformally flat spacetime following the Einstein's field equation and admits  CI is shear-free, and irrotational. Its energy density is constant over the space-like hypersurface orthogonal to the four-velocity vector.
	\end{theorem}
	\begin{proof}
		U. C. De and others stated that a result in \cite{MallickUCD}  as  \textquotedblleft A general relativistic spacetime with Codazzi type of energy-momentum tensor  is shear-free, irrotational, and its energy density is constant over the space-like hypersurface orthogonal to the four-velocity vector \textquotedblright. From equation \eqref{2.34n}, it is clear that the energy-momentum tensor of this spacetime is Codazzi type. By using this statement, the proof is clear. 
	\end{proof}
	\begin{theorem}
		A conformally flat spacetime, which satisfies  Einstein's field equations and admits CI, is a representation of inflation.
		
	\end{theorem}
	\begin{proof}
		Theorem \ref{Th2.8}  states that  $p + \mu$ = 0. This implies  that the fluid spacetime behaves as  spacetime of  cosmological constant \cite{KarmerD}.   According to \cite{Mazumdar}, this is also referred to as the $\bf Phantom~Barrier$.
	\end{proof}
	\begin{remark}
		All the Theorems from Theorem \ref{Th2.4} to Theorem \ref{Th2.8n} are also  valid for conformally flat spacetime that obeys the  Einstein's field equations without a cosmological constant and admits curvature inheritance.
	\end{remark}
	\begin{theorem}
		If a conformally flat spacetime admits CI and satisfies Einstein's field equations, then the velocity vector field U is infinitesimally spatially isotropic with respect  to the unit time-like vector field U.
	\end{theorem}
	\begin{proof}
		From equations \eqref{2.8} and \eqref{2.21}	we have 
		\begin{equation}\label{3.9n}
			R^h_{ijh} = \frac{R}{12} (g_{hh} g_{ij} - g_{hj} g_{ih}).
		\end{equation}
		According to Karcher \cite{Karcher}, a Lorentzian manifold is infinitesimally spatially isotropic with respect to a time-like unit vector field if its curvature tensor R satisfies the relation,
		\begin{equation}\label{3.10}
			(g(X,Z)Y - g(Y,Z)X) K = 	R(X,Y)Z,
		\end{equation}  
		where  $R(X,U)U= mX$ for  X$\in U^{\perp}$ and for all X, Y, Z $\in U^{\perp}$, where K and m are real valued function on the spacetime. From equations  \eqref{3.9n} and \eqref{3.10}   we get 
		\begin{equation}
			K  = \frac{R}{12} ~~~~   \mbox{and}  ~~~~  m = - \frac{R}{12}.
		\end{equation}
		Thus, the proof is completed.
	\end{proof}
	\begin{theorem}
		The sectional curvature determined by two vectors $X, Y \in $ $U^{\perp} $  and vectors X and U are equivalent, if  a conformally flat spacetime  admits  CI with perfect fluid.
	\end{theorem}
	\begin{proof}
		Let $X, Y \in $ $U^{\perp},$ $\kappa_{1}$ and $\kappa_{2}$ represents the sectional curvature of $(V_4, g)$ defined by X, Y and  X, U respectively. Then
		\begin{equation}
			\kappa_{1} = \frac{g(R(X,Y)Y,X)}{g(X,X) g(Y,Y) - (g(X,Y))^{2}} = \frac{R}{12},
		\end{equation}  
		and 
		\begin{equation}
			\kappa_{2} = \frac{g(R(X,U)U,X)}{g(X,X) g(U,U) - (g(X,U))^{2}} = \frac{R}{12}.
		\end{equation} 
		The proof is completed.
	\end{proof}
	
	\begin{theorem}
		Let  $(V_4, g)$ be a perfect fluid spacetime which is also conformally flat   satisfying Einstein's field equations without a cosmological constant does not admit curvature inheritance symmetry for a purely electromagnetic distribution.
	\end{theorem}
	
	\begin{proof}
		Considering the purely electromagnetic distribution of Einstein's field equations
		\begin{equation}\label{2.44n}
			R_{ij} = \kappa T_{ij},
		\end{equation}
		where $\kappa$ is gravitational constant. Let us assume that a conformally flat spacetimes admits curvature inheritance. From Theorem \ref{Th2.2} and equation \eqref{2.44n} yields
		\begin{equation}\label{2.45n}
			\frac{R}{4} g_{ij} = \kappa T_{ij}.
		\end{equation}
		\noindent From equations \eqref{2.2}, \eqref{2.45}  and \eqref{2.45n},  we get
		\begin{equation}
			R = 4 \kappa p.
		\end{equation}
		\noindent On the other hand by using equation \eqref{2.45n} yields
		\begin{equation}\label{2.47n}
			\kappa T = R.
		\end{equation} 
		\noindent If $T = 0$ (purely electromagnetic distribution), then  $R = 0$. Since a conformally flat spacetime is an Einstein, form equation \eqref{2.8}, it is flat. Thus, the proof is completed.
	\end{proof}
	
	\section{Example of de Sitter spacetime.}\label{sec3}
	Using the coordinates $t$, $r$, $\theta$, and $\phi$ as provided below, let $R^{4}$ be equipped with the following metric:
	\begin{equation}\label{3.1n}
		ds^{2} =   a^{2}(t) [(sin^{2}\theta d\phi^{2} + d\theta ^{2}) r^{2} + \frac{dr^{2}}{1 - \kappa r^{2}}] - dt^{2}.
	\end{equation}
	This metric is appropriate for a homogenous and isotropic universe and is known as the Robertson-walker metric. We assume that $\kappa$ = 0 in equation \eqref{3.1n}. This metric was first introduced by Alexander Friedemann on 1922 (\cite{Weinberg}- \cite{Liddle}) after solving Einstein's field equations under some assumptions for the contents of the universe in equation \eqref{3.1n},  comoving radial coordinates are denoted by $r$, comoving cosmic time by $t$, scale factor by $a(t)$, and metric on a unit 2-sphere by $d \varOmega^{2} $ (where $d \varOmega^{2}  = d\theta ^{2} + sin^{2} \theta d\phi^{2} )$, $\kappa$ = 1, 0, -1.
	
	The de-Sitter metric is equivalent to $a(t) = e^{Ht}$ and $\kappa$ = 0,  where $H$ is constant, so that in this case Friedmann-Lemaitre- Robertson-Walker metric ($\kappa$ = 0)
	\begin{equation}\label{3.2n}
		ds^{2} =  - dt^{2} + a^{2}(t) (dx^{2} + dy^{2} + dz^{2}).
	\end{equation} 
	
	\noindent For the case of the Robertson-Walker metric
	\begin{equation}\label{3.3n}
		R_{ij} = 0, ~~~~~ (i \neq j).
	\end{equation}
	This means that the  only non-zero elements of the Ricci tensor are only $R_{00}$, $R_{11}$, $R_{22}$ and $R_{33}$.
	For $R_{00}$ we have 
	\begin{equation}\label{3.4n}
		R_{00} = - 3 \frac{\ddot{a}}{a},
	\end{equation}
	and for the $R_{ii}$ we have 
	\begin{equation}\label{3.5n}
		R_{ii} = g_{ii} (\frac{\ddot{a}}{a} + 2 (\frac{\dot{a}}{a})^{2}), ~~~~ i = 1,2,3.
	\end{equation}
	It is easy now to calculate the Ricci scalar by contracting the Ricci tensor, the value for scalar curvature 
	\begin{equation}\label{3.6n}
		R = 6 (\frac{\ddot{a}}{a} +  (\frac{\dot{a}}{a})^{2}).
	\end{equation}
	Since this spacetime is Einstein, then from equations \eqref{3.5n} and \eqref{3.6n}  we obtained
	\begin{equation}\label{3.7n}
		R_{ii} = \frac{3}{2}  (\frac{\ddot{a}}{a} + 2 (\frac{\dot{a}}{a})^{2}) g_{ii}.
	\end{equation}
	By using the equation \eqref{3.4n}  get $ a \ddot{a} = \dot{a}^{2}$, Solving this differential equation we get 
	\begin{equation}\label{3.8n}
		a = C_{1} e^{C_{2}t}, ~~~~~ (C_{1},  C_{2} >0).
	\end{equation}
	If we choose the value of $C_{1} = 1$,  we have
	\begin{equation}\label{3.9nn}
		a(t) = e^{C_{2} t}.
	\end{equation}
	From equation \eqref{3.9nn}, equation \eqref{3.2n} reduces to form
	\begin{equation}\label{3.10n}
		ds^{2} =  - dt^{2} + e^{C_{2} t} (dx^{2} + dy^{2} + dz^{2}).
	\end{equation}
	In this case, by the aid of equations \eqref{3.10n}, \eqref{3.4n}, \eqref{3.5n} and \eqref{3.6n} takes the following form 
	\begin{equation}\label{3.11n}
		R_{00} = - \frac{3}{4},  ~~~~ R_{11} = R_{22} = R_{33}  = \frac{3}{4} e^{C_{2} t}, R = 3.
	\end{equation}
	It is found that components of the conformal curvature tensor vanishes. According to \cite{K.L.DuggalandR.Sharma}, it is seen that the components of this vector field, where the line element in equation \eqref{3.10n} accepts a proper conformal killing vector field, are in the form of $(e^{C_{2}t},0,0,0)$ and the conformal function is in the form of $C_{2}e^{C_{2}t}$. If the calculation is made using the components of the curvature tensor and the Lie derivative of this vector field is calculated. It can be seen that this spacetime has conformal inheritance and Ricci conformal inheritance vectors.
	For this reason, this example is also important in terms of realizing the first Theorem. Considering Theorem \ref{Th2.8} for spacetime that does not contain cosmological constants and remembering that energy density is positive, the scalar curvature of spacetime.  It is clear that de Sitter is also positive. Therefore, this spacetime is also de Sitter spacetime. Then, we have next result,
	\begin{theorem}
		The spacetime given by equation \eqref{3.2n} of the Line element is a de Sitter spacetime with a conformal inheritance vector.
	\end{theorem}
	\begin{remark}
		If the  warping function $f=e^{C_{2}  t}$ with $1$-dimensional base $M_1$ and $3$-dimensional fibre $M_2$ then the warped product $M_1\times_f M_2$, is mentioned in the metrics equation \eqref{3.10n}.
	\end{remark}

	\section{Discussion}\label{sec12}
	
	In this work, we have established key results of a curvature inheritance symmetry where studying conformally flat spacetime in general relativity. A specific instance of the general solution $\hbar_{ij}  = 2 \phi g_{ij} + \lambda K_{ij}$   is the motivation derived from the equation  \eqref{3.8} for the vector $\xi$, where $ K_{ij}$ is a symmetric tensor of second order. The curvature identity of the CI symmetry is given by the expression  $ g_{ih} R^h_{jkl} + g_{jh} R^h_{ikl} = 0$. In this article, we assert that  CI symmetry will help to improve earlier results on the geometry and physics of conformally flat spacetime. To clarify our claim, we first showed that our Theorem \ref{Th2.1} is applicable to a different type of spacetime. Recall that from \cite{TupperHall}, the existence of covariant constant scalar curvature must preclude some spacetime, including the situation of conformally flat spacetime, from the physical application of our Theorem  \ref{Th2.2}.
	
	If  fluid spacetime  satisfies the equation $\mu + p = 0$ then  spacetime equivalent to  spacetime  of  cosmological constant  \cite{KarmerD}. According to Mazumdar \cite{Mazumdar}, this is also referred to as the phantom barrier. The same result obtained   in cosmology, known as inflation, which  describes the rapid expansion of spacetime \cite{Arslan}.
	
	The fact that the collection of CKVs constitutes a finite-dimensional Lie algebra structure is widely known. Contrary to this, smooth curvature symmetry vectors' Lie algebraic structure is not always finite-dimensional. We lose the Lie algebra structure of CIVs if the differentiability requirement is less than smooth \cite{hall}. In order to distinguish precisely between the several categories of spaces that have a finite-dimensional Lie algebra, a link between metric and curvature symmetries is thus recommended. The metrics of the conformally flat generalised plane wave spacetime are of the form \cite{TupperHall}
	\begin{equation}\label{5.2}
		ds^2 = - A(u) (y^2 + z^2) du^2 - 2 du dv + dy^2 +dz^2.
	\end{equation}
	Lie algebra of finite dimension  \cite{Hallshabbir} formed by the collection of all CKV is indicated by the symbol $\mathcal{C}$.   The algebra of \cite{AA Coley}  CKV on $V_4$ can have a maximum dimension of 15, which is possible if $V_4$ is conformally flat. These spacetimes support homothetic Lie algebra in seven dimensions. It follows that there are often eight CKVs that are all appropriate since the spacetime of the form admits at most minuscule a seven-dimensional inherited algebra. As a result, according to \cite{TupperHall}, the conformally flat generalised plane wave spacetime has an inherited algebra with a maximum dimension of eight. We deduce that the set of all CIV in conformally flat forms is a finite-dimensional Lie algebra  from Theorem \ref{Th2.1}.\\

	\section{Acknowledgments}
	
	The authors thank Professor Graham Hall, FRSE, of the Institute of Mathematics at the University of Aberdeen in Scotland, UK, for his insightful comments and valuable discussions that enhanced this paper. Additionally, we acknowledge IUCAA, Pune for providing the necessary facilities and research environment to complete this work.\\

\end{document}